\def\b{\beta}
\def\n{\nu}
\def\m{\mu}

\def\ra{\rightarrow}
\def\t{\tilde}

\def\G{{\rm GeV}}

\def\eV{{\rm eV}}
\def\sL{{\scriptscriptstyle L }}
\def\sM{{\scriptscriptstyle M }}
\def\sN{{\scriptscriptstyle N }}

\def\sR{{\scriptscriptstyle R }}

\def\sS{{\scriptscriptstyle S }}

\def\sX{{\scriptscriptstyle X }}

\def\diag{{\scriptscriptstyle\rm diag}}

\def\>{{\stackrel{>}{\scriptstyle\sim}}}
\def\<{{\stackrel{<}{\scriptstyle\sim}}}

\def\MNS{{\rm \sM\sN\sS}}
\documentstyle [11pt,epsf]{article}
\makeatletter

\@addtoreset{equation}{section}
\makeatother
\begin{document}
\baselineskip=15 pt
\setcounter{page}{1}
\thispagestyle{empty}
\topskip 2.5  cm
\topskip 0.5  cm
%\begin{flushright}
%{\normalsize 2001/12/21}
%\end{flushright}
\vspace*{1.0 cm}
\centerline{\LARGE \bf Lepton Flavor Violating Processes in} 
\vskip 0.5 cm
\centerline{\LARGE \bf Bi-maximal Texture of Neutrino Mixings} 
\vskip 1.5 cm
\centerline{\Large \bf 
Atsushi Kageyama\renewcommand{\thefootnote}{\fnsymbol{footnote}}
\footnote[1]{E-mail address: atsushi@muse.sc.niigata-u.ac.jp}
\ \ \ \ 
Satoru Kaneko\renewcommand{\thefootnote}{\fnsymbol{footnote}}
\footnote[2]{E-mail address: kaneko@muse.sc.niigata-u.ac.jp}
}
\vskip 0.3 cm
\centerline{\Large \bf 
Noriyuki Shimoyama\renewcommand{\thefootnote}{\fnsymbol{footnote}}
\footnote[3]{E-mail address: simoyama@muse.sc.niigata-u.ac.jp}
\ \ \ \ 
Morimitsu Tanimoto\renewcommand{\thefootnote}{\fnsymbol{footnote}}
\footnote[4]{E-mail address: tanimoto@muse.sc.niigata-u.ac.jp}
}
\vskip 0.5 cm
 \centerline{\it{Department of Physics, Niigata University, 
 Ikarashi 2-8050, 950-2181 Niigata, JAPAN}}
\vskip 1.7  cm
%\centerline{\bf ABSTRACT}\par
%\vskip 0.3 cm

{
We investigate the lepton flavor violation in the framework of the MSSM with
right-handed neutrinos taking the large mixing angle MSW solution in the 
quasi-degenerate and  the inverse-hierarchical neutrino masses.
We predict the branching ratio of $\mu \rightarrow
e+\gamma$ 
%and $\tau \rightarrow \mu+\gamma$ 
processes assuming the degenerate
right-handed Majorana neutrino masses.  We find that the branching ratio
in the quasi-degenerate neutrino mass spectrum is 100 times smaller
than the ones in the inverse-hierarchical 
and the hierarchical neutrino spectra.
We emphasize  that  the magnitude of  $U_{e3}$ is one of  important
ingredients to predict BR($\mu \rightarrow e +\gamma $).
The effect  of the deviation from the complete-degenerate
 right-handed Majorana neutrino masses are also estimated.
%Furtheremore, we examine  the $S_{3\sL}\times S_{3\sR}$ model, which gives
%the quasi-degenerate neutrino masses,  and the Shafi-Tavartkiladze model,
%which gives the inverse-hierarchical neutrino masses.
%Both predicted branching ratios of $\mu\rightarrow e+\gamma$ 
%are smaller than the experimantal bound.
}
%%%%%%%%%%%%%%%%%%%%%%%%%%%%%%%%%%%%%%%%%%%%%%%%%%%%%%%%%%%%%%%%%%%%%%%%%%%%%%
\newpage
%%%%%%%%%%%%%%%%%%%%%%%%%%%%%%%%%%%%%%%%%%%%%%%%%%%%%%%%%%%%%%%%%%%%%%%%%%%%%%%
\topskip 0. cm
%%%%%%%%% section 1 %%%%%%%%%%%
\section{Introduction}
% Super-Kamiokande  has almost confirmed  the neutrino oscillation
% in the atmospheric neutrinos, which favors the $\n_\mu\Ar \nu_\tau$
%process  \cite{SKam}.
%For the solar neutrinos \cite{SKamsolar,SNO}, the recent data of 
%the Super-Kamiokande and the SNO also suggest the neutrino oscillation
%$\n_e\Ar \nu_{\rm x}$ with the large mixing angle (LMA) MSW solution,  
%although other solutions are still allowed \cite{MSW}.
%If we take the LMA-MSW solution,  neutrinos are  massive and 
%the flavor mixings are almost  bi-maximal  in the lepton sector. 
% 
If neutrinos are massive and mixed in the SM, 
%there exists a source of the lepton flavor violation (LFV) through the
%off-diagonal elements of the neutrino Yukawa coupling matrix. 
%However, 
due to the smallness of the neutrino masses, 
the predicted branching ratios for 
the lepton flavor violation (LFV) 
are so tiny that they 
are completely unobservable.
On the other hand,
in the supersymmetric framework  the situation is quite different.
Many authors have already studied the LFV  in the minimal supersymmetric standard model (MSSM) with right-handed neutrinos assuming the relevant neutrino 
mass matrix \cite{LFV1,LFV2,Sato,Casas}.
In the MSSM with soft breaking terms, 
there exist  lepton flavor violating terms such as 
off-diagonal elements of slepton mass matrices 
$\left({\bf m}^2_{\tilde \sL}\right)_{ij}$,
$({\bf m}_{ \t e_\sR}^2)_{ij}$ 
and trilinear couplings ${\bf A}^{e}_{ij}$.
Strong bounds on these matrix elements come from requiring branching ratios 
for LFV processes to be below observed ratios.  
For the present, the most stringent bound comes from  the 
$\mu \rightarrow e +\gamma $ decay 
(${\rm BR}(\m\ra e+\gamma)<1.2\times 10^{-11}$) \cite{MuEGamma}.
However, if the LFV  occurs at tree level in the soft breaking terms, 
the branching ratio of this process exceeds the experimental bound
considerably.
Therefore  one assumes that the LFV does not occur at tree level 
in the soft parameters. 
This is realized by taking the assumption that soft parameters 
such as $\left({\bf m}^2_{\t \sL}\right)_{ij}$,
$({\bf m}_{\t e_\sR}^2 )_{ij}$,  ${\bf A}^{e}_{ij}$,
are universal {\it i.e.}, proportional to the unit matrix. 
%This assumption follows from the minimal supergravity (m-SUGRA).  
However, even though there is no flavor violation at tree level, 
it is generated by the effect of
the renormalization group equations (RGE's) via neutrino Yukawa couplings. 
Suppose that  neutrino masses are produced by the  see-saw mechanism 
\cite{seesaw}, there are  the right-handed neutrinos 
above a  scale $M_{\sR}$.
Then neutrinos 
have  the Yukawa coupling matrix ${\bf Y}_\nu$  with  off-diagonal entries
in the basis of the diagonal charged-lepton Yukawa couplings.
 The  off-diagonal elements of ${\bf Y}_\nu$ drive off-diagonal ones 
in the $\left({\bf m}_{\rm\t \sL}^2\right)_{ij}$ and 
${\bf A}^{e}_{ij}$ matrices through the RGE's running \cite{Borzumati}.

%%%%%%%%%%%%%%%%%%%%%%%%%%%%%%%%%%%%%%%%%%%%%%%%%%%%%%
One can construct  ${\bf Y}_\nu$ by the recent data of neutrino oscillations.
   Assuming that oscillations need only accounting for 
 the solar and the atmospheric neutrino data, we take  the 
LMA-MSW solution of the solar neutrino.  
Then, the lepton mixing matrix,
 which may be called  the MNS matrix or the MNSP  matrix \cite{MNS,Po}, 
is given in ref.\cite{FT}. 
Since the data of neutrino oscillations only indicate
the differences of the mass square   $\Delta m^2_{ij}$,
  neutrinos have three possible mass spectra:  the hierarchical spectrum
  $m_{\n 3}\gg  m_{\n 2} \gg  m_{\n 1}$ , the quasi-degenerate one
  $m_{\n 1}\simeq m_{\n 2}\simeq m_{\n 3}$ 
  and the  inverse-hierarchical one  $m_{\n 1}\simeq m_{\n 2} \gg  m_{\n 3}$.
\section{LFV in the MSSM with Right-handed Neutrinos}
%\subsection{Yukawa Couplings}
%In this section, 
%we introduce the general expression of the neutrino Yukawa coupling 
%${\bf Y}_\n$, which is useful in the following arguments, 
%and investigate the LFV triggered by the neutrino Yukawa couplings.
The superpotential of the lepton sector is described as follows:
\begin{eqnarray}
 W_{\rm lepton}
&=&
{\bf Y}_e LH_de_{\sR}^c+{\bf Y_\n} LH_u\n_{\sR}^c
+\frac{1}{2}{\n_\sR^c}^T {\bf M_\sR}\n_\sR^c \ ,
\end{eqnarray}
where $H_u, H_d$ are chiral superfields for Higgs doublets, $L$ is
the left-handed lepton doublet, $e_\sR$ and $\n_\sR$ are the right-handed
charged lepton 
and the neutrino superfields, respectively. The ${\bf Y}_e$ 
is the Yukawa coupling matrix for the charged lepton,
${\bf M_\sR}$ is Majorana mass matrix of the right-handed neutrinos. We take
${\bf Y}_e$ and ${\bf M_\sR}$ to be diagonal.

It is well-known that the neutrino mass matrix is given as 
%\begin{eqnarray}
$
 {\bf m_{\n}}=
\left({\bf Y_\n} v_u\right)^{\rm T}{\bf M}_{\bf \sR}^{-1}
\left({\bf Y_\n} v_u\right),
$
%\label{eqn:mn}
%\end{eqnarray}
via the see-saw mechanism, where $v_u$ is the vacuum
expectation value (VEV) of Higgs $H_u$. 
%In eq.(\ref{eqn:mn}), the Majorana mass term for left-handed neutrinos is not
%included since we consider the minimal extension of the MSSM.
The neutrino mass matrix ${\bf m_\n}$ is diagonalized by a single unitary
 matrix
$
{\bf m_\n^{\diag}}\equiv{\bf U}_{\MNS}^{\rm T}{\bf m_\n} {\bf U_{\MNS}} 
$, 
where ${\bf U_{\MNS}}$ is the lepton mixing matrix. 
Following the expression in ref.\cite{Casas}, we write the neutrino Yukawa
 coupling as 
\begin{eqnarray}
{\bf Y_\n}
=
\frac{1}{v_u}
\sqrt{{\bf M_\sR^{\diag}}}\ {\bf R}\ \sqrt{{\bf m_\n^{\diag}}} \ 
{\bf U}^{\rm T}_{\MNS} \ ,
\label{YEW}
\end{eqnarray}
%or explicitly
%\begin{equation}
%{\bf Y_\nu}
%= \frac{1}{v_u}  \left (\matrix{\sqrt{M_{\sR 1}}& 0 & 0\cr
%  0 & \sqrt{M_{\sR 2}}  & 0 \cr  0 & 0 & \sqrt{M_{\sR 3}} \cr  } \right) 
%  {\bf R}
%  \left (\matrix{\sqrt{m_{\n 1}}& 0 & 0\cr
%  0 & \sqrt{m_{\n 2}}  & 0 \cr  0 & 0 & \sqrt{m_{\n 3}} \cr  } \right)  
% {\bf U}^{\rm T}_{\MNS}\ ,
%\label{YEW}
%\end{equation}
%\noindent 
where $\bf R$ is a $3\times 3$ orthogonal matrix, which depends on models.
%Details are given in Appendix A.

%%%%%%%%%%%%%%%%%%%%%%%%%%%%%%%%%%%%%%%%%%%%%%%%%%%%%%%%%%%%%%

At first, let us take the degenerate right-handed Majorana
masses  $M_{\sR1}=M_{\sR2}=M_{\sR3}\equiv M_R$.  
This assumption is reasonable for the case of the quasi-degenerate
neutrino masses.
Otherwise a big conspiracy would be needed between ${\bf Y_\nu}$ 
and ${\bf M}_{\bf\sR}$.  
This assumption is also taken for cases of the inverse-hierarchical and
the hierarchical neutrino masses.
We also discuss later the effect of the deviation from the degenerate 
right-handed Majonara neutrino masses.
Then, we get
\begin{eqnarray}
% {\bf Y_\nu}
%= \frac{\sqrt{M_{\sR}}}{v_u} \  {\bf R} \ 
%  \left (\matrix{\sqrt{m_{\n 1}}& 0 & 0\cr
%  0 & \sqrt{m_{\n 2}}  & 0 \cr  0 & 0 & \sqrt{m_{\n 3}} \cr  } \right)  
%{\bf U}^{\rm T}_{\MNS}\ , 
%\label{yukawa}
%\end{eqnarray}
%\noindent and
%\begin{eqnarray}
 {\bf Y_\n^{\dagger}Y_\n}
&=&
\frac{M_\sR}{v_u^2}
{\bf U_{\MNS}}
\left (
\matrix{
 m_{\n 1}& 0 & 0\cr
 0 & m_{\n 2}& 0\cr 
 0 & 0 &m_{\n 3}\cr
}
\right)  
{\bf U}_{\MNS}^{\rm T}\ .
\label{yy}
\end{eqnarray}
%or equivalently,
%\begin{eqnarray}
% \left({\bf Y_\n^{\dagger}Y_\n}\right)_{\alpha\beta}
%&=&
%\frac{M_\sR}{v_u^2}\sum_{i=1}^{3}m_{\n i}U_{\alpha i}U_{\beta i}^*\ ,
%\end{eqnarray} 
%where $U_{\alpha\beta}$'s are the elements of ${\bf U_{\MNS}}$.
It is remarked that $\bf Y_\n^{\dagger}Y_\n$ is independent of
 $\bf R$ in the case of  $M_{\sR1}=M_{\sR2}=M_{\sR3}\equiv M_\sR$.
As mentioned in the previous section, there are three possible neutrino
mass spectra.
The hierarchical type ($m_{\n 1}\ll m_{\n 2}\ll m_{\n 3}$) 
gives the neutrino mass spectrum as
\begin{eqnarray}
 m_{\n 1}\sim 0 \  ,\qquad m_{\n 2}=\sqrt{\Delta m^2_{\odot}} \  ,
  \qquad m_{\n 3}=\sqrt{\Delta m^2_{\rm atm}}\ ,
\end{eqnarray}
the quasi-degenerate type ($m_{\n 1}\sim m_{\n 2}\sim m_{\n 3}$) gives 
\begin{equation}
 m_{\n 1}\equiv m_\nu \ ,  
 \qquad  m_{\n 2}=m_\nu + \frac{1}{2 m_\nu}\Delta m^2_{\odot}\  , \qquad
 m_{\n 3}=m_\nu +  \frac{1}{2 m_\nu} \Delta m_{\rm atm}^2 \ , 
\label{eqn:degeneratetype}
\end{equation}
and the inverse-hierarchical type ($m_{\n 1}\sim m_{\n 2}\gg m_{\n 3}$) gives
\begin{equation}
 \quad  m_{\n 2}
\equiv
\sqrt{\Delta m^2_{\rm atm}} \  , \quad
 m_{\n 1}
=
 m_{\n 2}-\frac{1}{2 m_{\n 2}}\Delta m_{\odot}^2 \ , \quad  
 m_{\n 3}\simeq 0\ .
\end{equation}
%%%%%%%%%%%%%%%%%%%%%%%%%%%%%%%%%%%%%%%%%%%%%
%%%%%%%%%%%%%%%%%%%%%%%%%%%%%%%%%%%%%%%%%%%%%
We take the typical values 
$\Delta m^2_{\rm atm}=  3\times  10^{-3} \eV^2$ and 
$\Delta m_{\odot}^2= 7 \times  10^{-5}  \eV^2$ 
in our calculation of the LFV.

We take the typical mixing angles of the LMA-MSW solution such as 
$s_{23}=1/\sqrt{2}$ and $s_{12}=0.6$ \cite{FT}, in which  
  the lepton mixing matrix is given in  terms of the standard
  parametrization of 
the mixing matrix \cite{PDG}. 
%as follows:
%\begin{equation}  
%{\bf U_{\MNS}}
%=  \left (\matrix{ c_{13} c_{12} & c_{13} s_{12} &  s_{13} e^{-i \phi}\cr 
%  -c_{23}s_{12}-s_{23}s_{13}c_{12}e^{i \phi} & 
%c_{23}c_{12}-s_{23}s_{13}s_{12}e^{i \phi} &   s_{23}c_{13} \cr
%  s_{23}s_{12}-c_{23}s_{13}c_{12}e^{i \phi} & 
% -s_{23}c_{12}-c_{23}s_{13}s_{12}e^{i \phi} &  c_{23}c_{13} \cr} \right ) \ ,
%\end{equation} 
%where  $s_{ij}\equiv \sin{\theta_{ij}}$ and 
% $c_{ij}\equiv \cos{\theta_{ij}}$ are mixings in vacuum, 
% and $\phi$ is the $CP$ violating phase.  
The reacter experiment of  CHOOZ  \cite{Chooz} presented a upper bound of 
$s_{13}$.  We use the  constraint from the two flavor analysis, which is 
$s_{13}\leq 0.2$ in our calculation.
If we take account of  the recent result of the three flavor analysis
\cite{Three},  the upper bound of $s_{13}$ may be smaller than 0.2.
Then,  if we use the results in \cite{Three}, our results  of $\mu \rightarrow e+\gamma$ are reduced at most by a factor of two. 
In our calculation, the  CP violating phase is neglected for simplicity.
%%%%%%%%%%%%%%%%%%%%%%%%%%%%%%%%%%%%%%%%%%%%%%%%%%

%%%%%%%%% section 2-2 %%%%%%%%%%%
%\subsection{LFV in Slepton Masses}

Since SUSY is spontaneously broken at the low energy, 
we consider the MSSM with the soft SUSY breaking terms:
\begin{eqnarray}
-{\cal{L}}_{\rm soft}
%&=&
%({\bf m}_{\tilde \sQ}^2)_{ij} {\tilde Q}_{i}^{\dagger}{\tilde Q}_{j}
%+({\bf m}_{\tilde u}^2)_{ij} {\tilde u}_{\sR i}^* {\tilde u}_{\sR j}
%+({\bf m}_{\tilde d}^2)_{ij} {\tilde d}_{\sR i}^* {\tilde d}_{\sR j}
%\nonumber \\
=
({\bf m}_{\tilde \sL}^2)_{ij} {\tilde L}_{i}^{\dagger}{\tilde L}_{j}
+({\bf m}_{\tilde e}^2)_{ij} {\tilde e}_{\sR i}^* {\tilde e}_{\sR j}
+({\bf m}_{\tilde \nu}^2)_{ij} {\tilde \nu}_{\sR i}^* {\tilde \nu}_{\sR j}
%& &+{\widetilde m}^2_{H_d}H_d^{\dagger} H_d
%+{\widetilde m}^2_{H_u}H_u^{\dagger} H_u
%+(B \mu H_d H_u
%+\frac{1}{2}B_{\nu ij} M_{\sR ij}{\tilde \nu}_{\sR i}^* {\tilde \nu}_{\sR j}^* + h.c.)
%\nonumber \\
%
%+ ( {\bf A}^{d}_{ij}  H_d {\tilde d}_{\sR i}^*{\tilde Q}_{j}
%   +{\bf A}^{u}_{ij}  H_u {\tilde u}_{\sR i}^*{\tilde Q}_{j}
   +{\bf A}^{e}_{ij}  H_d {\tilde e}_{\sR i}^*{\tilde L}_{j}
   +{\bf A}^{\nu}_ {ij}H_u {\tilde \nu}_{\sR i}^*{\tilde L}_{j}\ ,
%
%\nonumber %\\
%& & 
%+\frac{1}{2}M_1 {\tilde B}_L^0 {\tilde B}_L^0
%+\frac{1}{2}M_2 {\tilde W}_L^a {\tilde W}_L^a
%+\frac{1}{2}M_3 {\tilde G}^a {\tilde G}^a +h.c.)\ ,
%
\end{eqnarray}
where $%{\bf m}_{\tilde Q}^2, {\bf m}_{\tilde u}^2, {\bf m}_{\tilde d}^2, 
{\bf m}_{\tilde \sL}^2, {\bf m}_{\tilde e}^2$ and ${\bf m}_{\tilde \nu}^2$
are mass-squares of %the left-handed squark, the right-handed up squark, 
%the right-handed down squark, 
the left-handed charged slepton, 
the right-handed charged slepton and the sneutrino, respectively. 
%The ${\widetilde m}^2_{H_d}$ and ${\widetilde m}^2_{H_u}$ are  
%mass-squares of Higgs,   
$
%{\bf A}_{d}, {\bf A}_{u}, 
{\bf A}_{e}$ and ${\bf A}_{\nu}$ are A-parameters.
% for squarks and 
%for sleptons.
%, and 
%$M_1, M_2$ and $M_3$ are the gaugino masses, 
%respectively.

%%%%%%%%%%%%%%%%%%%%%%%%%%%%%%%%%%%%%%%%%%%%%%%%% 
Note that the lepton flavor violating processes  come from 
diagrams including  non-zero off-diagonal elements of the soft
parameter.
In this paper we assume the m-SUGRA, therefore we put the assumption 
of universality for soft SUSY breaking terms at the unification scale:
\begin{eqnarray}
({\bf m}_{\tilde \sL}^2)_{ij}
=
({\bf m}_{\tilde e}^2)_{ij}
=({\bf m}_{\tilde \nu}^2)_{ij}
=\cdots
=\delta_{ij}m_{0}^2
\ \ ,\ \ 
%{\widetilde m_{H_d}}^2
%&=&
%{\widetilde m_{H_u}}^2
%=m_0^2
%\ ,
%\nonumber\\
{\bf A}^{\nu}
=
{\bf Y}_{\nu}a_0m_0,~
{\bf A}^{e}
={\bf Y}_{e}a_0m_0
\ ,
%\nonumber\\
%{\bf A}^{u}
%&=&
%{\bf Y}_{u}a_0m_0,~
%{\bf A}^{d}
%={\bf Y}_{d}a_0m_0
%\ ,
\end{eqnarray}
where  $m_0$ and $a_0$ stand for the universal scalar mass and the universal 
A-parameter, respectively.
%Because of the universality, the LFV is not caused
%at the unification scale. 

%
%%%%%%%%%%%%%%%%%%%%%%%%%%%%%%%%%%%%%%%%%%%%%%%%%%%%%%%%%
%To estimate the soft parameters at the low energy,
%we need to know the effect of radiative corrections.
%As a result, the lepton flavor conservation is violated
%at the low energy.

The RGE's  for the left-handed slepton soft mass are given by  
\begin{eqnarray}
&&\mu \frac{d}{d \mu}({\bf m}_{\tilde \sL}^2)_{ij}
=
\left.\mu \frac{d}{d \mu}({\bf m}_{\tilde \sL}^2)_{ij}\right|_{\rm MSSM}
\nonumber\\
&&\ \ +
\frac{1}{16 \pi^2}
\left[
({\bf m}_{\tilde \sL}^2 {\bf Y}_{\nu}^{\dagger}{\bf Y}_{\nu}
+{\bf Y}_{\nu}^{\dagger}{\bf Y}_{\nu}{\bf m}_{\tilde \sL}^2)_{ij}
+
2({\bf Y}_{\nu}^{\dagger}{\bf m}_{\tilde \nu}{\bf Y}_{\nu}
+{\tilde m}_{H _{u}}^2 {\bf Y}_{\nu}^{\dagger}{\bf Y}_{\nu}
+{\bf A}_{\nu}^{\dagger}{\bf A}_{\nu})_{ij}
\right],\ \ 
\end{eqnarray}
while  the first term in the right hand side 
is the normal MSSM term which has no LFV, 
and the second one is a source of the LFV through the off-diagonal elements 
of  neutrino Yukawa couplings. 
%The RGE's are summarized in Appendix B.

%%%%%%%%% section 3 %%%%%%%%%%%                                  
\section{Numerical Analyses of Branching Ratios}

Let us calculate the branching ratio of $e_{i}\rightarrow e_{j}+\gamma\
(j<i)$.
%The amplitude of this process is given as
%\begin{eqnarray}
% T=e\epsilon^{\alpha*}(q)\bar{u}_{j}(p)m_{e_{i}}i\sigma_{\alpha\beta}q^{\beta}
%(A^{\sL}P_{\sL}+A^{\sR}P_{\sR})u_{i}(q-p)\ ,
%\label{amplitude}
%\end{eqnarray}
%where $u_{i}$ is the wave function of $i$-th charged lepton $e_{i}$,  
%$p$ and $q$ are momenta of $e_{j}$ and photon, respectively, 
%$e$ is the electric charge, 
%$\epsilon$ is the polarization vector of photon, and  
%$P_{\sL,\sR}$ are  projection operators : $P_{\sL,\sR}=(1\mp\gamma_{5})/2$. 
%The $A^{\sL,\sR}$ are decay amplitudes and explicit forms are given in Appendix C.
%It is easy to see that this process changes chirality of the charged lepton.
The decay rate can be calculated using $A^{\sL,\sR}$ as 
\begin{eqnarray}
\Gamma(e_{i}\rightarrow e_{j}+\gamma)=
\frac{e^{2}}{16\pi}m_{e_{i}}^{5}
(|A^{\sL}|^{2}+|A^{\sR}|^{2})\ .
\label{rate}
\end{eqnarray}
Since we know the relation $m_{e_{i}}^2\gg m_{e_{j}}^2$, then we can
expect $|A^{\sR}|\gg |A^{\sL}|$.
%The $A^{\sL,\sR}$ contain the contribution of the neutralino loop and
%the chargino loop as seen in fig.1. 
%%%%%%%%%%%%%%%  Figure 1  %%%%%%%%%%%%%%%%%
%\begin{figure}
%\epsfxsize=10.0 cm
%\centerline{\epsfbox{diagram.ai}}
%\caption{Feynman diagrams which contribute to the branching ratio of
% $e_{i}\rightarrow e_{j}+\gamma$. There are two types of diagrams,
% (a)\,neutralino-slepton loop and (b)\,chargino-sneutrino loop.}
%\end{figure}
%We calculate the branching ratio using (\ref{rate}) and the formulas in
%Appendix C. 
%In order to clarify parameter dependence, 
%let us present an approximate estimation. 
The decay amplitude is approximated as 
\begin{eqnarray}
 | A^{\sR}|^2 \simeq
\frac{\alpha_{2}^{2}}{16\pi^{2}}
\frac{|(\bigtriangleup {\bf m}_{\tilde{\sL}}^{2})_{ij}|^{2}}{m_{\sS}^{8}}\tan^{2}\beta \ ,
\end{eqnarray}
where  $\alpha_{2}$ is  the gauge coupling constant of $SU(2)_{L}$ and  
$m_{S}$ is a SUSY particle mass.
The RGE's develop the off-diagonal elements of the slepton mass matrix 
and A-term:
%These terms at the low energy are approximated as 
\begin{eqnarray}
(\bigtriangleup {\bf m}_{\tilde{\sL}}^{2})_{ij}
\simeq
-\frac{(6+2a_{0}^2)m_{0}^2}{16\pi^2}({\bf Y_{\nu}^{\dag}Y_{\nu}})_{ij}
\ln
\frac{M_{\sX}}{M_{\sR}}\ ,
\end{eqnarray}
where $M_{\sX}$ is the GUT scale. 
%Therefore, off-diagonal elements of ${\bf (Y_{\nu}^{\dag}Y_{\nu}})_{ij}$ are
%the crucial quantity to estimate the branching ratio.

%As discussed in section 2, 
%${\bf (Y_{\nu}^{\dag}Y_{\nu})}_{ij}$ is given by neutrino masses and mixings
%at the electroweak scale.
%Therefore, we can compare the quantity ${\bf (Y_{\nu}^{\dag}Y_{\nu})}_{ij}$ 
%among the cases of  three neutrino mass spectra: 
%the degenerate, the inverse-hierarchical and the hierarchical masses. 
%In this section, 
%we present numerical results in these three cases. 
%%Here, we use eq.(\ref{rate}) and the vertex functions in Appendix C 
%%for the calculation of the branching ratio including the RGE's effect.
%
%
%%%%%%%%%% section 3-1-1 %%%%%%%%%%%        
%\subsection{$\mu \rightarrow e+\gamma$}

We present a qualitative discussion
on $({\bf Y}_{\nu}^{\dag}{\bf Y}_{\nu})_{21}$
before predicting  the branching ratio BR$(\mu \rightarrow e+ \gamma)$.   
This is given in terms of neutrino masses and mixings at the electroweak scale
as follows:
\begin{equation}
({\bf Y}_{\nu}^{\dag}{\bf Y}_{\nu})_{21}=
\frac{M_{\sR}}{v_{u}^{2}}
\left[U_{\mu2}U_{e2}^{*}(m_{\nu 2}-m_{\nu 1})+
 U_{\mu3}U_{e3}^{*}(m_{\nu 3}-m_{\nu 1})
\right] \ ,
\label{normal}
\end{equation}
\noindent 
where $v_u\equiv v\sin\b$ with $v=174\G$ is taken  
as an usual notation and  the unitarity condition of the lepton mixing matrix
elements is used.
Taking the three cases of the neutrino mass spectra,
%: the degenerate, 
%the inverse-hierarchical  and the normal hierarchical masses
one obtains the following forms, repectively,
\begin{eqnarray}
({\bf Y}_{\nu}^{\dag}{\bf Y}_{\nu})_{21}
\simeq &&
\frac{M_{\sR}}{\sqrt{2}v_{u}^{2}}\frac{\Delta m^2_{\rm atm}}{2m_\nu}
\left[
\frac{1}{\sqrt{2}}U_{e2}^* \frac{\Delta m_{\odot}^2}{\Delta m^2_{\rm atm}}
+ U_{e3}^*
\right] \ ,          
\quad ({\rm Degenerate}) 
\nonumber \\
\nonumber \\
\simeq  &&
\frac{M_{\sR}}{\sqrt{2}v_{u}^{2}}\sqrt{\Delta m_{\rm atm}^2}
\left[ 
\frac{1}{2\sqrt{2}} U_{e2}^*\frac{\Delta m_{\odot}^2}{\Delta m^2_{\rm atm}} 
- U_{e3}^*
\right] \ , 
\quad ({\rm Inverse }) 
\label{muemass3} \\
\nonumber \\
\simeq  &&
\frac{M_{\sR}}{\sqrt{2}v_{u}^{2}}\sqrt{\Delta m_{\rm atm}^2}
\left[
\frac{1}{\sqrt{2}}U_{e2}^*
\sqrt{\frac{\Delta m_{\odot}^2}{\Delta m^2_{\rm atm}}}
+ U_{e3}^*
\right] \ , 
\quad ({\rm Hierarchy }) \nonumber
\end{eqnarray}
\noindent
where we take the maximal mixing for the atmospheric neutrinos. 
Since  $U_{e2}\simeq 1/\sqrt{2}$ for the bi-maximal mixing matrix, 
the first terms in the square brackets of the right hand sides of  
eqs.(\ref{muemass3}) can be estimated by putting the experimental data.
For the case of the degenerate neutrino masses,
$({\bf Y}_{\nu}^{\dag}{\bf Y}_{\nu})_{21}$ depends on the unknown neutrino mass scale 
$m_{\nu}$.  
As one takes the smaller $m_\nu$, one predicts the larger branching ratio. 
In our calculation, we take $m_{\nu}=0.3 \eV$
%%%%%%%%%%%%%%%%%%%%%%%%%%%%%%%%%%%%%%%%%%%%%%%%%%%%
%\footnote{Recently, a positive observation of the neutrinoless double beta 
%decay was reported in \cite{data}, where the degenerate neutrino mass of 
% $m_{\nu}=0.3 \eV$ is a typical one.}
%%%%%%%%%%%%%%%%%%%%%%%%%%%%%%%%%%%%%%%%%%%%%%%%%%%%
, which is close to the upper 
bound from the neutrinoless double beta decay experiment \cite{Beta},
and also leads to the smallest branching ratio. 

%%%%%%%%%%%%%%%%%%%%%%%%%%%%%%%%%%%%%%%%%%%%%%%%5
We also note that the degenerate case  gives 
the smallest branching ratio  BR$(\mu \rightarrow e +\gamma)$ among the
three cases as seen in  eqs.(\ref{muemass3}) owing to the scale of
$m_{\nu}$.
It is easy to see the fact that the second terms in eqs.(\ref{muemass3}) 
are dominant as far as  
$U_{e3} ~\>~ 0.01({\rm degenerate}), ~ 0.01({\rm inverse})$ 
and $0.07({\rm hierarchy})$, respectively.
The magnitude and the phase of $U_{e3}$ are important in the comparison between cases of the inverse-hierarchical and the normal hierarchical masses.
In the limit of  $U_{e3}=0$, the predicted branching ratio in the 
case of the normal hierarchical masses is larger than the other
one. However, for $U_{e3}\simeq 0.2$ 
the predicted branching ratios are almost the same in both cases.

 At first, we  present numerical results in the case of the degenerate 
neutrino masses assuming ${\bf M_\sR}=M_\sR {\bf 1}$.
%%%%%%%%%%%%%%%%%%%%%%%%%%%%%%%%%%%%%%
%%%%%%%%%%%%%%%%%%%%%%%%%%%%%%%%%%%%%%
%%%%%%%%%%%%%%%%%%%%%%%%%%%%%%%%%%%%%%
The magnitude of $M_\sR$ is constrained considerably if we impose the
  $b-\tau$  unfication of Yukawa couplings  \cite{btau}.
In the case of $\tan\beta \leq 30$, the lower bound of
$M_\sR$ is approximately $10^{12}\G$.  We take also $M_\sR \leq 10^{14}\G$,
in order that neutrino Yukawa couplings remain below ${\cal O}(1)$.
 Therefore, we use  $M_\sR = 10^{12},\ 10^{14}\G$ 
 in our following calculation.

%%%%%%%%%%%%%%%%%%%%%%%%%%%%%%%%%%%%%%%
%%%%%%%%%%%%%%%%%%%%%%%%%%%%%%%%%%%%%%%
%%%%%%%%%%%%%%%%%%%%%%%%%%%%%%%%%%%%%%%
We take a universal scalar mass $(m_0)$ for all scalars and
$a_0=0$ as a universal A-term at the GUT scale ($M_\sX=2\times 10^{16}$ GeV).
The branching ratio of  $\mu \rightarrow e+\gamma$ is given 
versus the left-handed 
selectron mass $m_{\tilde e_\sL}$ for each $\tan\b=3,\  10,\  30$
and a fixed wino mass $M_2$ at the electroweak scale.
In fig.\ref{D1402}, the branching ratios
are  shown for  $M_2=150,\  300\ \G$
 in the case of $U_{e3}=0.2$ with $M_R=10^{14} \G$ and $m_{\nu}=0.3$eV,
in which the solid curves correspond to  $M_2=150 \G$
and the dashed ones to  $M_2=300 \G$.
The threshold of the selectron mass is determined by the
recent LEP2 data \cite{LEP} for   $M_2=150 \G$, however, for $M_2=300
\G$,  determined by the  constraint that  the left-handed slepton  should be 
 heavier than the neutralinos.
As the  $\tan\b$ increases, the branching ratio increases
because the decay amplitude from the SUSY diagrams is approximately 
proportional to  $\tan\b$ \cite{LFV1}.
It is found that the branching ratio is  almost larger than the
experimental upper bound in the case of   $M_2=150 \G$.
On the other hand, the predicted values are smaller than the
experimental bound except for $\tan\b=30$  in the case of $M_2=300 \G$.

 Our predictions depend on $M_\sR$ strongly, because the magnitude of
the neutrino Yukawa coupling is determined by  $M_\sR$
as seen in eq.(\ref{YEW}).
If  $M_\sR$ reduces to  $10^{12} \G$,
the branching ratio becomes $10^{4}$ times smaller since it
is proportional to $M_\sR^2$.
The numerical result is shown in fig.\ref{D1202}.
%%%%%%%%%%%%%%%%%%%%%%%%%%%%%%%%%%%%%%%%%%%%%%%%%%%
%We will examine a model \cite{FTY,O3}, which  
%gives the degenerate neutrino masses with 
%$U_{e3}\sim 0.05$ in section 4. 
%%%%%%%%%%%%%%%%%%%%%%%%%%%%%%%%%%%%%%%%%%%%%%%%%%%%%

 Next we show results in the case of the inverse-hierarchical neutrino 
masses.  As expected in eq.(\ref{muemass3}), the branching ratio
is much larger than the one in the degenerate case.
 In fig.\ref{I1402},
the branching ratio is shown for  $M_2=150,\  300\ \G$
 in the case of $U_{e3}=0.2$ with $M_\sR=10^{14} \G$.
 In fig.\ref{I14005}, the branching ratio is shown for 
$U_{e3}=0.05$ with  $M_\sR=10^{14} \G$.
The  $M_R$ dependence is the same as the case of 
 the quasi-degenerate neutrino masses.
The predictions almost exceed the experimental bound  as far as
$U_{e3}\geq 0.05$, $\tan\beta\geq 10$ and  $M_\sR\simeq 10^{14} \G$.
This result is based on the assumption  ${\bf M}_{\bf \sR}=M_{\sR}{\bf 1}$,
however, it is not  guaranteed in the case of the inverse-hierarchical 
neutrino masses.
%We will examine  a typical  model  \cite{Shafi},  which gives 
%${\bf M}_{\bf \sR} \not =M_{\sR}{\bf 1}$ in section 4.
%%%%%%%%%%%%%%%%%%%%%%%%%%%%%%%%%%%%%%%%%%%%%%%%%%%%%%%%%%%%%

For comparison,
we show the branching ratio in the case of the hierarchical neutrino 
masses in fig.\ref{H1402}. 
It is similar to the case of the inverse-hierarchical neutrino masses.
The branching ratio in the case of the 
degenerate neutrino masses is $10^2$ times smaller than 
the one in the inverse-hierarchical 
and the hierarchical neutrino spectra.

%%%%%%%%%%%%%%%%%%%%%%%%%%%%%%%%%% non-zero a %%%%%%%%%%%%%%%%%%%
%In our numerical analyses we assumed $a_{0}=0$ at the GUT scale
%$M_{\sX}$ for simplicity.
%Let us comment on the A-term dependence, namely $a_{0} \neq 0$ at $M_{\sX}$.
%We estimate the branching ratio for $a_{0}= \pm 1$ at $M_{\sX}$ 
%$({\bf A}={\bf Y}a_{0}m_{0})$.
%In the degenerate type, the predicted branching ratio is 
%1.02($a_{0}=1$), 1.07($a_{0}=-1$) times as large as the one 
%in the case of $a_{0}=0$ (tan$\beta=30$, $U_{e3}=0.2$). 
%In the inverse-hierarchical type, the predicted branching ratios are
%1.56($a_{0}=1$), 1.54($a_{0}=-1$) times as large as the one 
%in the case of $a_{0}=0$ (tan$\beta=30$, $U_{e3}=0.2$). 
%Therefore the A-term dependence is insignificant in our analyses.
%
%%%%%%%%%%%%%%  m=0 %%%%%%%%%%%%%%%%
%In our calculations,
%we use the universality condition at $M_{\sX}$.
%We also examine the no-scale condition $m_{0}=0$ at $M_{\sX}$. 
%It is found that the predicted branching ratio is $10$ times smaller 
%than the one in the case of non-zero universal scalar mass.

%%%%%%%%%%%%%%%%%%%%%%%%%%%%%%%%%%%%%%%%%%%%%%%%%
%\subsection{Non-degeneracy Effect of ${\bf M}_{R}$}

The analyses in the previous section depend on the
  assumption of $M_{\sR1}=M_{\sR2}=M_{\sR3}\equiv M_R$.
In the case of the quasi-degenarate neutrino masses 
in eq.(\ref{eqn:degeneratetype}) 
this complete degeneracy of ${\bf M}_{\sR}$ may be deviated in the 
following magnitude without fine-tuning:
\begin{eqnarray}
\frac{M_{\sR3}^2}{M_{\sR1}^2} \simeq 1 \pm  \frac{\Delta m _{\rm atm}^2}{m_\nu^2}\ ,
\qquad 
\frac{M_{\sR2}^2}{M_{\sR1}^2} \simeq 1 \pm  \frac{\Delta m_{\odot}^2}{m_\nu^2} \ .
\end{eqnarray}
\noindent
%Therefore, we  parametrize  $M_{\sR}$ as 
%\begin{eqnarray}
%{\bf M_{\sR}}
%=
%M_{\sR}
%\left(
%\begin{array}{ccc}
%1&       0       &0 \\
%0 &1+\varepsilon_{2}&0 \\
%0 &       0       & 1+\varepsilon_{3}
%\end{array}
%\right) \ ,
%\end{eqnarray}
%\noindent
%where $\varepsilon_2\simeq  \Delta m_{\odot}^2/2 m_\nu^2$ and 
%$\varepsilon_3\simeq  \Delta m _{\rm atm}^2/2 m_\nu^2$.
By using eq.(\ref{YEW}), we obtain
%\begin{equation}
%{\bf Y}^{\dagger}_{\nu}{\bf Y}_{\nu}
%=
%\frac{M_{\sR}}{v^2_{u}}
%{\bf U}_{\MNS}
%\left(
%\begin{array}{ccc}
%\sqrt{m_{\nu1}}&   0            & 0\\
%        0       &\sqrt{m_{\nu2}}& 0\\
%         0      &           0    & \sqrt{m_{\nu3}}
%\end{array}
%\right)
%{\bf K}
%\left(
%\begin{array}{ccc}
%\sqrt{m_{\nu1}}  &    0             & 0\\
%      0           & \sqrt{m_{\nu2}} &0 \\
%        0         &           0      & \sqrt{m_{\nu3}}
%\end{array}
%\right)
%{\bf U}^{\rm T}_{\MNS} \ ,
%\end{equation}
%\noindent where
%\begin{equation}
%{\bf K}\equiv  {\bf R}^{\dagger}
%\left(
%\begin{array}{ccc}
%1  &    0             & 0\\
% 0  & 1+\varepsilon_{2}  & 0\\
%0   &             0    & 1+\varepsilon_{3}
%\end{array}
%\right)
%{\bf R} \ .
%\end{equation}
%\noindent 
%Then, we have
%\begin{eqnarray}
%({\bf Y}^{\dagger}_{\nu}{\bf Y}_{\nu})_{21}
%=
%\frac{M_{\sR}}{v^2_{u}}
%\sum^{3}_{i,j}
%U_{2 i}U_{1 j}
%(K_{ij}\sqrt{m_{\nu i}}\sqrt{m_{\nu j}}) \ ,
%\end{eqnarray}
%\noindent with
%\begin{eqnarray}
%K_{ij}
%=
%\delta_{ij}
%+\varepsilon_{2}R_{2i}R_{2j}
%+\varepsilon_{3}R_{3i}R_{3j} \ ,
%\end{eqnarray}
%where we used ${\bf R}^{\rm T}{\bf R=1}$
%%\footnote{we assume ${\bf R}$ to be real for simplicity. }.
%Then, we get 
\begin{eqnarray}
({\bf Y}^{\dagger}_{\nu}{\bf Y}_{\nu})_{21}
=
\left.
({\bf Y}^{\dagger}_{\nu}{\bf Y}_{\nu})_{21}
\right |_{\bf M_{\sR \propto 1}}+
\Delta({\bf Y}^{\dagger}_{\nu}{\bf Y}_{\nu})_{21} \ ,
\end{eqnarray}
where the first term is the  $({\bf Y}^{\dagger}_{\nu}{\bf Y}_{\nu})_{21}$ element
in eq.(\ref{normal}), which corresponds to the  ${\bf M_{\sR \propto 1}}$,
while the second term stands for the deviation from it as follows:
\begin{equation}
\Delta({\bf Y}^{\dagger}_{\nu}{\bf Y}_{\nu})_{21}=
\frac{M_{\sR}}{v^2_{u}}
\sum^{3}_{i,j}
U_{2 i}U_{1 j}
\sqrt{m_{\nu i}}\sqrt{m_{\nu j}}
(\varepsilon_{2}R_{2i}R_{2j}
+\varepsilon_{3}R_{3i}R_{3j}) \  .
\end{equation}
In order to estimate the second term, we use
$\varepsilon_2=0.0001$ and $\varepsilon_3=0.01$ taking account of
 $\varepsilon_2\simeq \Delta m_{\odot}^2/2 m_\nu^2$ and 
$\varepsilon_3\simeq  \Delta m _{\rm atm}^2/2 m_\nu^2$, where
$m_\nu=0.3 \eV$ is put.
Since $m_{\nu i}\simeq m_{\nu j}$  and $R_{ij} \leq 1$,
we get 
\begin{eqnarray}
\Delta({\bf Y}^{\dagger}_{\nu}{\bf Y}_{\nu})_{21}
%&\sim&
%\frac{M_{\sR}}{v^2_{u}}
%\sum^{3}_{i,j}
%U_{2 i}U_{1 j}
%m_{\nu}
%\varepsilon_{3}R_{3i}R_{3j}
%\nonumber\\
\leq
%\frac{M_{\sR}}{v^2_{u}}
%\frac{1}{2\sqrt{2}}
%m_{\nu}
%\varepsilon_{3}
%\sim 
3.5 \times 10^{-3} \ .
\end{eqnarray}
Taking this maximal value, we can estimate the branching ratio 
as follows:
\begin{eqnarray}
\frac{BR({\rm non\mbox{-}degenerate}\ M_{\sR})}{BR({\rm degenerate} \ M_{\sR})}
\leq
\left(\frac{2.6 + 3.5}{2.6} \right)^2 \simeq  5.5 \ .
\end{eqnarray}
Therefore, the enhancement due to the second term is at most 
factor 5.
This conclusion does not depend on the specific form of ${\bf R}$

Consider the case of the  inverse-hierarchical type of neutrino masses.
We take $\varepsilon_{2}\sim 0.01$ 
with the similar argument of the quasi-degenerate type neutrino masses,
because 
$m_{\nu 1}$ and $m_{\nu 2}$ are almost degenerate and 
$\varepsilon_2\simeq  \Delta m _{\odot}^2/2 \Delta m _{\rm atm}^2$ 
in this case.
Then, we get  
\begin{eqnarray}
\Delta({\bf Y}^{\dagger}_{\nu}{\bf Y}_{\nu})_{21}
%&=&
%\frac{M_{\sR}}{v^2_{u}}
%\sum^{2}_{i,j}
%U_{2 i}U_{1 j}
%m_{\nu1}
%\varepsilon_{2}R_{2i}R_{2j}
%\nonumber\\
\leq
%\frac{M_{\sR}}{v^2_{u}}
%\frac{1}{2\sqrt{2}}
%m_{\nu1}
%\varepsilon_{2}
%\sim  
0.063 \times 10^{-2} \ ,
\end{eqnarray}
where we assume $\varepsilon_{2} \geq \varepsilon_{3}$ and  use
$m_{\nu3}\simeq 0$, $m_{\nu1} \simeq  m_{\nu2} \simeq 0.054$eV  
and $R_{ij}\leq 1$.
Taking  the maximal value, we get 
\begin{eqnarray}
\frac{BR({\rm non\mbox{-}degenerate} \ M_{\sR})}{BR({\rm degenerate} \ M_{\sR})}
\leq 
\left(
\frac{2.7 + 0.063}{2.7}
\right)^2
\simeq 1.04 \ .
\end{eqnarray}
%%%%%%%%%%%%%%%%%%%%%%%%%%%%%%%
Thus, the effect of the $\Delta({\bf Y}^{\dagger}_{\nu}{\bf Y}_{\nu})_{21}$ is very small
in the case of the inverse hierarchical neutrino masses.

\newpage
\begin{figure}
\twocolumn
\epsfxsize=6.0 cm
\centerline{\epsfbox{D1402.ai}} 
\caption{
%Predicted branching ratio 
BR$(\mu \rightarrow e+\gamma)$
%versus the left-handed selectron mass for $\tan\b=3,\ 10,\ 30$
in the case of the degenerate neutrino masses.
%Here $M_\sR=10^{14}\G$ and $U_{e3}=0.2$ are taken.
%The solid curves correspond to  $M_2=150 \G$ and the dashed ones to  
%$M_2=300 \G$.
%A horizontal dotted line denotes the experimental upper bound.
}
\label{D1402}
\epsfxsize=6.0 cm
\centerline{\epsfbox{D1202.ai}}
\caption{
%Predicted branching ratio 
BR$(\mu \rightarrow e+\gamma)$
%versus the left-handed selectron mass for $\tan\b=3,\ 10,\ 30$
in the case of the case of degenerate neutrino masses.}
%Here $M_\sR=10^{12}\G$ and $U_{e3}=0.2$ are taken.
%The solid curves correspond to  $M_2=150 \G$ and the dashed ones to  
%$M_2=300 \G$.}
\label{D1202}
%%%%%%%%%%%%%%%%  Figure 4  %%%%%%%%%%%%%%%%
\epsfxsize=6.0 cm
\centerline{\epsfbox{I1402.ai}}
\caption{
%Predicted branching ratio 
BR$(\mu \rightarrow e+\gamma)$
%versus the left-handed selectron mass for $\tan\b=3,\ 10,\ 30$
in the case of the inverse-hierarchical neutrino masses.}
%
%Here $M_\sR=10^{14}\G$ and $U_{e3}=0.2$ are taken.
%The solid curves correspond to  $M_2=150 \G$ and the dashed ones to  
%$M_2=300 \G$.}
\label{I1402}
\end{figure}
\begin{figure}
%%%%%%%%%%%%%%%%  Figure 5  %%%%%%%%%%%%%%%%%
\epsfxsize=6.0 cm
\centerline{\epsfbox{I14005.ai}}
\caption{
%Predicted branching ratio 
BR$(\mu \rightarrow e+\gamma)$
%versus the left-handed selectron mass for $\tan\b=3,\ 10,\ 30$
in the case of the inverse-hierarchical neutrino masses.
%Here $M_\sR=10^{14}\G$ and $U_{e3}=0.05$ are taken.
%The solid curves correspond to  $M_2=150 \G$ and the dashed ones to  
%$M_2=300 \G$.
}
\label{I14005}
\end{figure}
%%%%%%%%%%%%%%%%  Figure 6  %%%%%%%%%%%%%%%%
\begin{figure}
\epsfxsize=6.0 cm
\centerline{\epsfbox{H1402.ai}}
\caption{
%Predicted branching ratio 
BR$(\mu \rightarrow e+\gamma)$
%versus the left-handed selectron mass for $\tan\b=3,\ 10,\ 30$
in the case of the hierarchical neutrino masses.
%Here $M_\sR=10^{14}\G$ and $U_{e3}=0.2$ are taken.
%The solid curves correspond to  $M_2=150 \G$ and the dashed ones to  
%$M_2=300 \G$.
}
\label{H1402}
\end{figure}

\end{document}